\documentclass[12pt]{article}

\usepackage{fancybox} 

\usepackage{a4}

\usepackage{amsmath}
\usepackage{amsfonts}

\newcommand{\ol}{\overline}
\newcommand{\wh}{\widehat}
\newcommand{\wt}{\widetilde}

\newcommand{\rap}[2]
{\setbox1=\hbox{#1}%
\setbox2=\hbox to\wd1{\hss #2\hss}%
\mbox{\rlap{\box1}\box2}}

\def\rot{\mathop{\rm rot}}
\def\tr{\mathrm{Tr}}

\def\half{{1\over2}}

\def\[{\left[}
\def\]{\right]}
\def\({\left(}
\def\){\right)}

\def\N{{\cal N}}

\def\S{{\cal S}}

\def\CC{{\mathbf C}}

\def\SSS{{\mathbf S}}
\def\RR{{\mathbf R}}
\def\ZZ{{\mathbf Z}}

\def \be {\begin{equation}}
\def \ee {\end{equation}}
\def \bea {\begin{eqnarray}}
\def \eea {\end{eqnarray}}

\author{Y.~Imamura and S.~Yokoyama}
\begin{document}

\begin{titlepage}
\title{
\begin{flushright}
\normalsize{ UT-10-11\\
TIT/HEP-605\\
Aug 2010}
\end{flushright}
       \vspace{2cm}
Twisted Sectors in Gravity Duals of $\N=4$ Chern-Simons Theories
       \vspace{2cm}}
\author{
Yosuke Imamura\thanks{E-mail: \tt imamura@phys.titech.ac.jp}$^{~1}$
and Shuichi Yokoyama\thanks{E-mail: \tt yokoyama@hep-th.phys.s.u-tokyo.ac.jp}$^{~2}$
\\[30pt]
{\it $^1$ Department of Physics, Tokyo Institute of Technology,}\\
{\it Tokyo 152-8551, Japan}\\
{\it $^2$ Department of Physics, University of Tokyo,}\\
{\it Tokyo 113-0033, Japan}
}
\date{}

\maketitle
\thispagestyle{empty}

\vspace{0cm}

\begin{abstract}
\normalsize
We study Kaluza-Klein modes of a $d=7$, $\N=2$ vector multiplet
in $AdS_4\times\SSS^3$.
Such modes arise in the context of AdS/CFT 
as dual objects of a class of gauge invariant operators
in $\N=4$ Chern-Simons theories.
We confirm that the Kaluza-Klein modes precisely
reproduce the BPS spectrum of the operators.
\end{abstract}

\end{titlepage}


\section{Introduction}
\label{intr}
Since the discovery of BLG model\cite{Bagger:2006sk,Bagger:2007jr,Bagger:2007vi,Gustavsson:2007vu,Gustavsson:2008dy} and ABJM model\cite{Aharony:2008ug},
quiver type supersymmetric Chern-Simons theories
have attracted a great interest.
They are expected to be realized on  M2-branes placed in
cone geometries,
and many checks have been done to confirm the duality
between Chern-Simons theories with different numbers of supersymmetries
and M-theory on the corresponding
AdS backgrounds.

A strong support for this duality is obtained by the analysis of indices.
For the $\N=6$ supersymmetric Chern-Simons theory,
ABJM model, a superconformal index which encodes the spectrum
of gauge invariant operators are computed in
\cite{Bhattacharya:2008bja,Kim:2009wb}, and it agrees with the
index obtained
by the Kaluza-Klein analysis in the dual geometry\cite{Bhattacharya:2008zy}.

In this paper, we focus on the gravity duals of
$\N=4$ supersymmetric Chern-Simons theories with unitary gauge groups\cite{Aharony:2008ug,Hosomichi:2008jd,Benna:2008zy,Imamura:2008nn}.
Even though the analysis of these theories on the gauge theory side
is parallel to that for ABJM model,
it is quite interesting because
there are new ingredients on the gravity side.
The $7$-dimensional internal space of
the dual geometry for an $\N=4$ Chern-Simons theory
is a certain orbifold of $7$-sphere.
In general, it includes orbifold singularities,
and the emergence of twisted sectors is expected.

An ${\cal N}=4$ Chern-Simons theory with unitary gauge group
has quiver type matter contents and is described by a circular quiver diagram.
There are two kinds of hypermultiplets in this theory:
untwisted and twisted hypermultiplets.
Each edge in the quiver diagram represents one of them.
If the number of edges is even, and
$m$ untwisted and $m$ twisted hypermultiplets appear
alternately in the diagram,
the theory is obtained by $\ZZ_m$ orbifolding of ABJM model.
This theory is similar to a four-dimensional ${\cal N}=2$ CFT
obtained as a $\ZZ_m$ orbifold of the ${\cal N}=4$ supersymmetric Yang-Mills
theory.
It is described by the $A_{m-1}$ type quiver diagram,
and has $\ZZ_m$ symmetry shifting the diagram.
The dual geometry of this theory also has a singular locus, and twisted sectors
arise.
The field-operator correspondence in the twisted sectors is
studied in \cite{Aharony:1998xz,Gukov:1998kk},
and it is shown that operators in
the twisted sectors are characterized by non-vanishing $\ZZ_m$ charges.

In the case of the ${\cal N}=4$ Chern-Simons theory
which is obtained as an orbifolded ABJM model,
the existence of twisted sectors was first found in \cite{Choi:2008za}.
They computed a certain index for the theory
and extract non-vanishing contribution of the twisted sectors from it
as a deviation from what is
obtained by a naive orbifold projection from the ABJM model.
They also show that operators in twisted sectors are
characterized in the same way as the four dimensional ${\cal N}=2$ CFTs
by non-vanishing charges with respect to
$\ZZ_m$ symmetry which rotates the quiver diagram
with circumference $2m$ by even steps.
(We cannot shift the diagram by odd steps because two kinds of hypermultiplets
appear alternatively in it.)

The analysis in \cite{Choi:2008za} is restricted in
the perturbative sector
which does not include monopole operators.
The inclusion of monopole operators is
accomplished in \cite{Imamura:2009hc} by following the method
used in  \cite{Kim:2009wb} for ABJM model.
We can again use the $\ZZ_m$ symmetry to
specify twisted sectors.
Let us define magnetic charges of an operator by
\begin{equation}
m_a=\frac{1}{2\pi}\oint\tr F_a,
\end{equation}
for each vertex labeled by $a=1,\ldots,2m$ in the quiver diagram.
If these charges are not invariant under the $\ZZ_m$ symmetry,
such a monopole operator belongs to a twisted sector.
In \cite{Imamura:2008ji,Imamura:2009ur}
such monopole operators are identified to M2-branes wrapped on
vanishing cycles at singular loci.

In addition to theories obtained
as orbifolds of ABJM model,
there exist more general ${\cal N}=4$ Chern-Simons theories
which
include arbitrary numbers of two kinds of hypermultiplets
appearing in an arbitrary order in the quiver diagram.
Although such a theory is not obtained from ABJM model by
the orbifolding procedure,
the corresponding M2-brane background is again an orbifold of the flat space.
It is possible to construct different theories corresponding to
the same background geometry.
They can be distinguished on the gravity side only by the discrete torsion
of the three-form potential.
Because wrapped M2-branes couple to this torsion,
we can use monopole operators, which correspond to wrapped M2-branes,
as probes to establish the relation between
discrete torsions on the gravity side and boundary theories which share
the same moduli space.
This analysis has been done in \cite{Imamura:2009hc},
and the relation\cite{Witten:2009xu}
between the discrete torsions and
the linking numbers, which is defined by
the structure of the quiver diagram, is reproduced.
See also \cite{Klebanov:2010tj,Benishti:2010jn,Gutierrez:2010bb} 
about the relation between wrapped M2-branes and monopole operators.

In the previous works \cite{Choi:2008za} and \cite{Imamura:2009hc},
the twisted sector contribution to
indices are computed only on the
gauge theory side.
To confirm the AdS/CFT correspondence, it is necessary
to reproduce the same indices from Kaluza-Klein modes of fields localized on
the singular loci,
but it has not yet been done.
The purpose of this paper is to fill this gap.

As will be explained in detail in the next section, the
worldvolume of a singular locus is
$AdS_4\times \SSS^3/\ZZ_m$,
and on the locus a $d=7$, ${\cal N}=2$ $SU(n)$ vector multiplet
lives, where $m$ and $n$ are integers related to the
structure of the Chern-Simons theory.
We need the Kaluza-Klein spectrum of this vector multiplet.
At the linearized level, we do not
have to take account of
the interactions, and we can treat an $SU(n)$ vector multiplet
as a set of $n^2-1$ $U(1)$ vector multiplets.
What we will actually do in the following sections is
to investigate
the Kaluza-Klein spectrum of a $U(1)$ vector multiplet
in $AdS_4\times \SSS^3$, the covering space
of a singular locus.
Once we obtain this spectrum,
we can easily obtain the spectrum of $SU(n)$ fields on the singular locus
by collecting all the contribution of $n^2-1$ vector multiplets,
and carrying out an appropriate $\ZZ_n$ orbifold projection.

For the Kaluza-Klein analysis, we need the equations of
motion for the component fields of the vector multiplet.
Because the background spacetime is curved, we should take account of the
couplings to the background curvature.
We determine the couplings by Noether procedure.
The detailed derivation is given in Appendix.

This paper is organized as follows.
In section \ref{gsym}, we explain the structure of
the dual geometry with focusing on its symmetry.
In section \ref{BPS},
we briefly review $1/2$ BPS representations of the superconformal group $OSp(4|4)$.
In section \ref{KK}, we carry out Kaluza-Klein analysis for scalar fields in the vector multiplet.
Combining this with the knowledge of
representation theory given in section \ref{BPS},
we can obtain the whole Kaluza-Klein spectrum.
By using this spectrum, 
we compute a character and an index for the vector multiplet in section \ref{CandI}, and confirm that they agree with
the previous results obtained on the gauge theory side.
Section \ref{discussions} is devoted to Summary and discussions.
In Appendix \ref{consts} we derive the $d=7$, ${\cal N}=2$
supersymmetric action of
a vector multiplet on the curved background $AdS_4\times \SSS^3$,
and a direct derivation of the whole 
Kaluza-Klein spectrum is given in Appendix \ref{kkanalysis}.

\section{Global symmetries}\label{gsym}
The R-symmetry of a three dimensional $\N=4$
supersymmetric Chern-Simons theory
is $SO(4)_R \simeq SU(2)_R \times SU(2)'_R$. 
Two kinds of hypermultiplets are distinguished by the action of
the R-symmetry on the component fields.
Scalar fields in an untwisted hypermultiplet
belong to an $SU(2)_R$ doublet,
while ones in a twisted hypermultiplet form an
$SU(2)'_R$ doublet.

It would be instructive to understand symmetries of
$\N=4$ Chern-Simons theories from the geometric point of view.
The moduli space of an $\N=4$ Chern-Simons theory is
the symmetric product of $N$ copies of a certain
Abelian orbifold of $\CC^4$\cite{Benna:2008zy,Imamura:2008nn,Terashima:2008ba},
and the theory describes $N$ M2-branes in this orbifold.
If the background is $\CC^4$ itself,
the system possesses $\N=8$ supersymmetry,
and the R-symmetry is $SO(8)$,
the rotation group of $\CC^4$.
To obtain an $\N=4$ theory,
we first split $\CC^4$ into two $\CC^2$.
Correspondingly, we define the subgroup
$SO(4)\times SO(4)'\subset SO(8)$
which rotates two $\CC^2$ separately.
Because each $SO(4)$ is a product of two $SU(2)$ factors,
we have in total four $SU(2)$ factors in this subgroup.
We denote these four factors as follows.
\begin{equation}
SO(4)=SU(2)_R\times SU(2)_F,\quad
SO(4)'=SU(2)'_R\times SU(2)'_F.
\label{twoso4}
\end{equation}
The supercharges of the $\N=8$ theory
belong to the spinor representation ${\bf 8}_s$ of
$SO(8)$, and its branching in the subgroup
$SU(2)_R\times SU(2)_F\times SU(2)'_R\times SU(2)'_F$ is
\begin{equation}
{\bf8}_s\rightarrow
({\bf2},{\bf1},{\bf2},{\bf1})\oplus({\bf1},{\bf2},{\bf1},{\bf2}).
\label{8sbranch}
\end{equation}
If we perform the orbifold projection
by using an appropriate Abelian discrete subgroup
of $SU(2)_F\times SU(2)_F'$,
the latter half on the right hand side
in (\ref{8sbranch}) is projected out,
and we are left with $\N=4$ supersymmetry.
The remaining four supercharges are transformed
as a vector of
\begin{equation}
SO(4)_R=SU(2)_R\times SU(2)_R'.
\end{equation}
This group is nothing but the R-symmetry mentioned at the beginning  of this
section.
Note that $SO(4)_R$ is not any of two
$SO(4)$s in (\ref{twoso4}).

The orbifolding breaks $SU(2)_F$ and $SU(2)_F'$ into
Abelian subgroups which commute with the
orbifold group.
We denote them by $U(1)_P\subset SU(2)_F$ and $U(1)_P'\subset SU(2)_F'$.

The explicit form of
the background geometry of M2-branes
for an ${\cal N}=4$ Chern-Simons theory is
\begin{equation}
((\CC^2/\ZZ_p)\times(\CC^2/\ZZ_q))/\ZZ_k,
\label{modsp}
\end{equation}
where $p$ and $q$ are the numbers of two kinds of hypermultiplets in the theory and
$k$ is the Chern-Simons coupling constant.
See \cite{Imamura:2008nn} for detail.
The orbifold group is generated by the transformations
\begin{eqnarray}
(z_1,z_2,z_3,z_4)&\rightarrow&
(e^{\frac{2\pi i}{p}}z_1,e^{-\frac{2\pi i}{p}}z_2,z_3,z_4),\nonumber\\
(z_1,z_2,z_3,z_4)&\rightarrow&
(z_1,z_2,e^{\frac{2\pi i}{q}}z_3,e^{-\frac{2\pi i}{q}}z_4),\nonumber\\
(z_1,z_2,z_3,z_4)&\rightarrow&
(e^{\frac{2\pi i}{pk}}z_1,e^{-\frac{2\pi i}{pk}}z_2,
e^{\frac{2\pi i}{qk}}z_3,e^{-\frac{2\pi i}{qk}}z_4),
\end{eqnarray}
where $z_1,\ldots,z_4$ are complex coordinates of $\CC^4$.
ABJM model is the special case with $p=q=1$.
The dual geometry of the theory
is the near horizon limit of the classical solution of
M2-branes in the orbifold background,
and is the product of AdS$_4$ and the orbifolded seven sphere 
\begin{equation}
\(\SSS^7/(\ZZ_p\times\ZZ_q)\)/\ZZ_k,
\label{int}
\end{equation}
which is defined from
(\ref{modsp}) by the restriction
$|z_1|^2+|z_2|^2+|z_3|^2+|z_4|^2=1$.

An important feature of this dual geometry is
that it includes orbifold singularities.
The $\ZZ_p$ orbifolding in (\ref{int}) generates $A_{p-1}$ type
singularities.
The continuous set of the fixed points
forms
the fixed locus with topology $\SSS^3/\ZZ_{qk}$.
Similarly, the fixed locus associated with $\ZZ_q$ is
$\SSS^3/\ZZ_{pk}$.
We refer to these two fixed loci as $\S$ and $\S'$, respectively.
On $\S$
there exists a $d=7$, ${\cal N}=2$ $SU(p)$ vector multiplet.
The Cartan part of this arises from the localized zero modes
of the supergravity fields, while the non-Cartan part
arises from M2-branes wrapping on vanishing $2$-cycles
at the singular locus.
We also have an $SU(q)$ vector multiplet localized at the other
singular locus $\S'$.
These two loci can be treated in a similar way,
and we mainly focus on the vector multiplet in $\S$.

As we mentioned above, $\S$ is $\ZZ_{qk}$ orbifold
of its covering space $\wt\S=\SSS^3$.
Hereafter,
we investigate a vector multiplet in $AdS_4\times\wt\S$.
Once we obtain the Kaluza-Klein spectrum in $\wt\S$,
we can easily derive the spectrum in $\S$ by an appropriate
projection.

Among global symmetries in
(\ref{twoso4}), only $SU(2)'_R$ and $SU(2)'_F$ act on
$\wt\S$ transitively,
while $SU(2)_R$ and $SU(2)_F$ do not move the locus.
This means that from the viewpoint of the seven-dimensional theory
on the locus,
$SU(2)_R$ and $SU(2)_F$ are internal symmetries
while the other two are
parts of the isometry group
\begin{equation}
Sp(4,\RR)\times SU(2)'_R\times SU(2)'_F,
\end{equation}
of the seven-dimensional spacetime.
The last factor in this group, $SU(2)'_F$, is broken to $U(1)'_P$
by the orbifolding, and the other part, $Sp(4,\RR)\times SU(2)'_R$,
is a part of the bosonic subgroup
$Sp(4,\RR)\times SU(2)_R\times SU(2)'_R$
of the superconformal group $OSp(4|4)$.

\section{$1/2$ BPS representations}\label{BPS}
In this section we briefly review $1/2$ BPS representations
of the superconformal group $OSp(4|4)$.
To describe highest weights of irreducible representations,
we use Cartan generators $D$ (dilatation) and $j$ (spin)
for the conformal group $Sp(4,\RR)$,
$T_3$ for $SU(2)_R$, and $T_3'$ for $SU(2)_R'$.

In section \ref{CandI}
we will compute a superconformal index
of the Kaluza-Klein modes.
For the definition of the index, we use
only two supercharges out of the four,
and the choice of the two
breaks the R-symmetry $SO(4)_R$ down to
$SO(2)\times SO(2)$.
We denote generators of these two $SO(2)$ symmetries by
$H_1$ and $H_2$.
They are related to $T_3$ and $T_3'$ by
\begin{eqnarray}
H_1=T_3+T'_3, \quad H_2=-T_3+T'_3.
\label{ht}
\end{eqnarray}
Following \cite{Dolan:2008vc}, we define four supercharges $Q^I$ and $\ol Q^I$ ($I=1,2$)
carrying the Cartan charges shown in Table \ref{4susy}.
\begin{table}[htb]
\caption{The Cartan charges of four supercharges are shown.}
\label{4susy}
\begin{center} 
\begin{tabular}{ccccccc}
\hline
\hline
& $D$ & $j$ & $H_1$ & $H_2$ & $T_3$ & $T_3'$ \\ 
\hline
$Q^1$     & $1/2$ & $\pm1/2$ & $+1$ & $0$ & $1/2$ & $1/2$\\
$\ol Q^1$ & $1/2$ & $\pm1/2$ & $-1$ & $0$ & $-1/2$ & $-1/2$ \\
$Q^2$     & $1/2$ & $\pm1/2$ & $0$ & $+1$ & $-1/2$ & $1/2$ \\
$\ol Q^2$ & $1/2$ & $\pm1/2$ & $0$ & $-1$ & $1/2$ & $-1/2$ \\
\hline
\end{tabular} 
\end{center}
\end{table}
We will use $Q^1$ to define the index,
and then only $SO(2)$ generated by $H_1$ plays a role of
R-symmetry.

As usual, we use highest weights to specify representations.
The highest weights $T_3$ and $T_3'$
of an $SO(4)_R$ representation
are non-negative,
and there is the following bound for $H_1$ and $H_2$.
\begin{equation}
H_1\geq|H_2|.
\label{hgeqh}
\end{equation}
Short representations we are interested in are
$1/2$ BPS representations which saturate this bound.
There are two series of such representations, which are called $(\N,B,\pm)$
in \cite{Dolan:2008vc}.
($\N$ in the first slot refers to the number of
supersymmetry, and in this paper it is always $4$.)
A representation in each series is
specified by one integer $n$,
and the components of its highest weight are
\begin{equation}
(D,j,H_1,H_2)=(\frac{n}{2},0,\frac{n}{2},\pm\frac{n}{2}),\quad
n=1,2,\ldots 
\label{hw}
\end{equation}
The last component $H_2$ is positive for $(4,B,+)$,
and negative for $(4,B,-)$.
We denote the $1/2$ BPS representation with the highest weight \eqref{hw} by $(4,B,\pm)_n$.
The highest weight states $|0\rangle_{(4,B,\pm)}$
satisfy
\begin{equation}
Q^1|0\rangle_{(4,B,+)}=Q^2|0\rangle_{(4,B,+)}=0,\quad
Q^1|0\rangle_{(4,B,-)}=\ol Q^2|0\rangle_{(4,B,-)}=0.
\end{equation}

The spectra of $(4,B,\pm)_n$ representations are given in
\cite{Dolan:2008vc}.
Each of them is decomposed into six irreducible representations of
the bosonic subgroup $Sp(4,\RR)\times SO(4)$.
The decomposition of $(4,B,+)_n$ is shown in Table \ref{cps}.
That of $(4,B,-)_n$ is obtained from this
by exchanging $T_3$ and $T_3'$, and correspondingly,
flipping the sign of $H_2$.

Corresponding to the two fixed loci $\S$ and $\S'$, both
$(4,B,+)$ and $(4,B,-)$
arise from the twisted sectors.
Because $T_3$ does not move $\S$, it is an internal charge
in the context of the field theory in $\S$,
while $T_3'$ is an orbital angular momentum.
This fact implies that the Kaluza-Klein modes
in $\S$ should be identified with $(4,B,+)$,
which can take an arbitrarily large $T_3'$.
Contrary, the Kaluza-Klein modes in the other locus $\S'$
belong to the other series of representations $(4,B,-)$.

\begin{table}[htb]
\caption{The spectrum of the $1/2$ BPS representation $(4,B,+)_n$.
The representation is decomposed into six irreducible representations
of the bosonic subgroup $Sp(4,\RR)\times SO(4)_R$.
The highest weights of these representations are given.
$\Delta$ is defined later in (\ref{deltadef}).}
\label{cps}
\begin{center}
\begin{tabular}{cccccccc}
\hline
\hline
    $D$   & $j$ & $T_3$ & $T_3'$ & $H_1$ & $H_2$ & $\Delta$ & range\\
\hline
$\frac{n}{2}$  & $0$ & $0$ & $\frac{n}{2}$   & $\frac{n}{2}$ & $\frac{n}{2}$ & $0$ & ($n\geq1$) \\
$\frac{n+1}{2}$ & $\frac{1}{2}$ & $\frac{1}{2}$ & $\frac{n-1}{2}$ & $\frac{n}{2}$ & $\frac{n}{2}-1$ & $0$ & ($n\geq1$)\\
$\frac{n+2}{2}$  & $0$ & $1$ & $\frac{n-2}{2}$  & $\frac{n}{2}$ & $\frac{n}{2}-2$& $1$ & ($n\geq1$) \\
$\frac{n+2}{2}$   & $1$ & $0$ & $\frac{n-2}{2}$ & $\frac{n}{2}-1$ & $\frac{n}{2}-1$ & $1$ & ($n\geq2$) \\
$\frac{n+3}{2}$ & $\frac{1}{2}$ & $\frac{1}{2}$ & $\frac{n-3}{2}$ & $\frac{n}{2}-1$ & $\frac{n}{2}-2$ & $2$ & ($n\geq3$) \\
$\frac{n+4}{2}$  & $0$ & $0$ & $\frac{n-4}{2}$ & $\frac{n}{2}-2$ & $\frac{n}{2}-2$  & $4$ & ($n\geq4$) \\
\hline
\end{tabular}
\end{center}
\end{table}

\section{Kaluza-Klein spectrum}
\label{KK}
Let us determine the
Kaluza-Klein spectrum of a $d=7$, ${\cal N}=2$ vector multiplet in $AdS_4\times \SSS^3$.
To carry out this analysis, we need the action of the vector multiplet in
the seven-dimensional curved background.
In general, supersymmetry requires the coupling of
the component fields to the background curvature.
We can determine the action order by order
with respect to the background curvature
by using the Noether procedure.
A vector multiplet consists of
a gauge field $A_M$, a symplectic Majorana spinor field $\lambda$,
and three real scalar fields $\phi_i$ ($i=1,2,3$).
If we identify the background spacetime with the singular locus
$AdS_4\times \S$, the R-symmetry of this theory is $SU(2)_R$
defined in section \ref{gsym}, and the isometry of $\SSS^3$ is $SU(2)_R'\times SU(2)_F'$.
$\lambda$ and $\phi_i$ belong to the $SU(2)_R$ doublet and the triplet, respectively.
The supersymmetric action is
\begin{eqnarray}
S&=&
\int d^7x\sqrt{-g}\Big[
-\frac{1}{4}F_{MN}F^{MN}-\frac{1}{2}\ol\lambda\Gamma^MD_M\lambda-\frac{1}{2}\partial_M\phi_i\partial^M\phi_i
\nonumber\\&&
+\frac{3}{8L}(\ol\lambda \Gamma \lambda)
+\frac{1}{L^2}\phi_i\phi_i
-\frac{3}{4L}\epsilon^{kmn}A_kF_{mn}\Big],
\label{stot}
\end{eqnarray}
where $L$ is the AdS radius.
See Appendix \ref{consts} for a detailed derivation of this action.
This is invariant under the supersymmetry transformation
\begin{equation}
\delta\phi_i=i(\ol\epsilon\sigma_i\lambda),\quad
\delta\lambda=-i\sigma_i\Gamma^M\epsilon\partial_M\phi_i
+\frac{1}{2}F_{MN}\Gamma^{MN}\epsilon
+\frac{i}{L}\phi_i\sigma_i\Gamma\epsilon,\quad
\delta A_M=-(\ol\epsilon\Gamma_M\lambda).
\label{susytr}
\end{equation}
There are mass terms for the fermion $\lambda$ and the scalars
$\phi_i$.
We also have the Chern-Simons coupling for the gauge field.
Note that the tachyonic scalar mass $m^2=-2/L^2$
satisfies the Breitenlohner-Freedman bound
$m^2\geq -9/(4L^2)$.
These terms are inevitable to obtain the Kaluza-Klein spectrum
consistent with the gauge invariant operators in the boundary theory.
In this section, we derive only the spectrum of the
scalar fields.
A direct derivation of the full Kaluza-Klein spectrum of
the other component fields is
provided in Appendix \ref{kkanalysis}.

The equation of motion of the scalar fields
derived from (\ref{stot}) is
\begin{equation}
\Delta\phi_i
+\frac{2}{L^2}\phi_i=0,
\label{scalareom}
\end{equation}
where the Laplacian is defined with the
background $AdS_4\times \SSS^3$ metric
\begin{equation}
ds_7^2=L^2\frac{-dx_0^2
+dx_1^2
+dx_2^2
+dr^2}{r^2}+(2L)^2d\Omega_3^2.
\end{equation}
The scalar function in $\SSS^3$ can be expanded
by scalar spherical harmonics $Y_{0,(s,s)}$.
The quantum number $s=0,1/2,1,\ldots$ is the orbital angular momentum
in $\SSS^3$.
A harmonic $Y_{0,(s,s)}$ belongs to the
$SO(4)$($=SU(2)_R'\times SU(2)_F'$) representation
with the highest weight $(s,s)$,
and it actually has $(2s+1)^2$ components forming
the representation.

On the three-sphere with radius $2L$, eigenvalue of the Laplacian
for the scalar harmonic is
\begin{equation}
\Delta_{\SSS^3}Y_{0,(s,s)}=-\frac{1}{(2L)^2}4s(s+1)Y_{0,(s,s)}.
\end{equation}
The Kaluza-Klein expansion of
a scalar field $\phi$ is
\begin{equation}
\phi=\sum_s f_{(s,s)}Y_{0,(s,s)},
\end{equation}
where $f_{(s,s)}$ are scalar fields on $AdS_4$.

For the purpose of computing the conformal dimension
of the corresponding operators, it is sufficient to assume that
$f_{(s,s)}$ depends only on the radial coordinate $r$ in AdS$_4$.
By substituting this into the equation of motion
(\ref{scalareom}), we obtain
\begin{equation}
r^4\frac{d}{dr}\frac{1}{r^2}\frac{d}{dr}f_{(s,s)}-(s^2+s-2)f_{(s,s)}=0.
\end{equation}
This has two independent solutions
\begin{equation}
f_{(s,s)}\propto r^{s+2},\quad
r^{-s+1}.
\end{equation}
When we discuss field-operator correspondence,
we need to take one of these two solutions which is
normalizable.
If $s$ is sufficiently large,
only the former is normalizable.
Here we choose the former for every $s$,
which corresponds to an operator with conformal dimension $D=s+2$.

These Kaluza-Klein modes of scalar fields must be a part of
BPS representations given in Table \ref{cps}.
Because the scalar fields form an $SU(2)_R$ triplet,
we identify these modes with the third line in the table.
By comparison of the conformal dimension
and the $SU(2)_R'$ representations, we can relate the quantum number $n$ to the
orbital spin $s$ by
\bea
n= 2s+2,\quad
s=0,\frac{1}{2},\ldots.
\label{ns}
\eea 
For every $s$, we have $2s+1$ superconformal multiplets,
which are transformed as the spin $s$ representation of the
flavor group $SU(2)'_F$.
The spectrum $V$ is
\begin{equation}
V=\bigoplus_{s=0,\frac{1}{2},\ldots}(4,B,+)_{2(s+1)}\otimes (\mbox{$\bf s$ of $SU(2)'_F$}).
\label{dec}
\end{equation}
This fact is also conformed by the direct derivation of
the full Kaluza-Klein spectrum
given in Appendix \ref{kkanalysis}.
The results are summarized in Table \ref{spec}.
\begin{table}[htb]
\caption{The Kaluza-Klein spectrum of a vector multiplet $(A_M,\lambda,\phi_i)$ in $AdS_4\times \SSS^3$.
For each mode shown in the table,
there exists the other mode with $D$ replaced by $3-D$,
which is not normalizable for large $s$.
}
\label{spec}
\begin{center}
\begin{tabular}{cccccc}
\hline
\hline
fields       & $SU(2)_J$ & $SU(2)_R$ & $SU(2)_R'$ & $SU(2)_F'$ & $D$ \\
\hline
$\phi_i$  & $0$ & $1$ & $s$   & $s$   & $s+2$ \\
$A_M$  & $0$ & $0$ & $s+1$   & $s$   & $s+1$ \\
       & $1$ & $0$ & $s$   & $s$   & $s+2$ \\
& $0$ & $0$ & $s-1$   & $s$   & $s+3$ \\
$\lambda$
          & $\frac{1}{2}$ & $\frac{1}{2}$ & $s+\frac{1}{2}$ & $s$   & $s+\frac{3}{2}$ \\
& $\frac{1}{2}$ & $\frac{1}{2}$ & $s-\frac{1}{2}$ & $s$   & $s+\frac{5}{2}$ \\
\hline
\end{tabular}
\end{center}
\end{table}

\section{A character and an index}
\label{CandI}
In the last section, 
by combining the Kaluza-Klein analysis for the scalar fields
and the knowledge of the representation theory,
we determined the Kaluza-Klein spectrum \eqref{dec}.
With this result, we can easily obtain 
a character for the vector multiplet.

We define the superconformal character
for a representation $R$ by
\bea
\chi_{R}
= \tr_{R} \biggl(s^{2D} x^{2j} y^{T_3} y'{}^{T_3'}\biggr)
= \tr_{R} \biggl(s^{2D} x^{2j} y_1^{H_1} y_2^{H_2}\biggr),
\eea
where $\tr_R$ means the trace over the representation $R$.
We used Cartan generators $(D, j, T_3, T_3')$ for the middle expression
and $(D, j, H_1, H_2)$ for the last one.
These two choices of the Cartan generators for $SO(4)_R$ are related by
\eqref{ht},
and correspondingly, 
the two sets of variables $(y, y')$ and $(y_1, y_2)$ are related by
\bea
y= {y_1\over y_2}, \quad y'= y_1 y_2.
\eea

As is shown in Table \ref{cps}, $(4,B,+)_n$ is decomposed into
six irreducible representations of the bosonic subgroup $Sp(4,\RR)\times SO(4)_R$.
The character is obtained by summing up those for the six representations.
\bea
\chi_{(4,B,+)_n}
&=&
\chi^{\rm conf}_{(\frac{n}{2},0)}(s^2,x^2)\chi_{{n\over 2}}(y')
+\chi^{\rm conf}_{(\frac{n+1}{2},\frac{1}{2})}(s^2,x^2) \chi_{\half}(y) \chi_{{n-1\over 2}}(y')
\nonumber\\&&
+\chi^{\rm conf}_{(\frac{n+2}{2},1)}(s^2,x^2)\chi_{{n-2\over 2}}(y')
+\chi^{\rm conf}_{(\frac{n+2}{2},0)}(s^2,x^2)\chi_1(y) \chi_{{n-2\over 2}}(y')
\nonumber\\&&
+\chi^{\rm conf}_{(\frac{n+3}{2},\frac{1}{2})}(s^2,x^2)\chi_{\half}\(y\) \chi_{{n-3\over 2}}(y')
+\chi^{\rm conf}_{(\frac{n+4}{2},0)}(s^2,x^2)\chi_{{n-4\over 2}}(y')
\label{chi-4bn}
\eea
where $\chi^{\rm conf}_{(D,j)}$ is the character
of the irreducible representation of
the conformal group with highest weight $(D,j)$,
\begin{equation}
\chi^{\rm conf}_{(D,j)}=\tr_{(D,j)}(s^{2D}x^{2j})
=\frac{s^{2D}\chi_{j}(x^2)}{(1-s^2 x^2)(1-s^2)(1-s^2 x^{-2})},
\end{equation}
and $\chi_{j}(t)$ is the $SU(2)$ character for the spin $j$ representation
\bea
\chi_{j}(t) = {t^{j}-t^{-j-1} \over 1-t^{-1}} = t^{j}+\cdots+t^{-j},\quad
(\mbox{$\chi_{j}(t)=0$ for $j<0$}).
\eea

We can derive the character for the Kaluza-Klein modes \eqref{dec}
by summing up $\chi_{(4,B,+)_n}$ with an appropriate weight.
To include the information of the flavor group $SU(2)_F'$, 
we introduce a variable $z'$ for the Cartan generator $P'$ of $SU(2)_F'$.
We normalize $P'$ in a different way from $T_3$ and $T_3'$
so that its eigenvalues are integers.
We define
\begin{equation}
\chi_V
=\tr_{V} \biggl(s^{2D} x^{2j} y^{T_3} y'{}^{T_3'}z'^{P'} \biggr),
\label{ch}
\end{equation}
and with (\ref{dec}), we obtain
\bea
\chi_V= \sum^\infty_{s=0}\chi_{(4,B,+)_{2s+2}}\chi_s(z'^2).
\eea

In general, it is difficult to
calculate a character directly on the gauge theory side
due to quantum corrections.
We can avoid this by choosing the arguments of the character so that
the contributions of two states connected by a supercharge Q have opposite
signs and cancel each other.
Such a character is called an index, and it can be used conveniently
to check the AdS/CFT correspondence.
Let us choose $Q=Q^1_{\downarrow}$.
In this case, 
$OSp(4|4)$ algebra tells us that two Cartan generators $D+j$ and $H_2$
commute with $Q$,
and the index
\bea
I_{R}&=& \tr_{R} \biggl((-)^F  x'^{2\Delta} x^{2(D+j)} y_2^{H_2}\biggr),
\eea
does not depend on $x'$,
where $F$ is the fermion number operator, and $\Delta$ is defined by
\bea
\Delta=\{Q,Q^\dagger\}=D-(j+H_1).
\label{deltadef}
\eea
The index is related to the character by
\begin{equation}
I_{R}= \tr_{R} \biggl((-)^F  x^{2(D+j)} y_2^{H_2}\biggr)
=\chi_{R}(s\rightarrow xx', x\rightarrow -\frac{x}{x'}, y_1\rightarrow\frac{1}{x'^2},y_2).
\label{ch-ind}
\end{equation}
where we used $(-)^{F}=(-)^{2j}$ to show the second equality. 
By substituting \eqref{chi-4bn} into the relation \eqref{ch-ind}, 
we obtain the superconformal index for a half BPS representation $(4,B,+)_n$ 
\bea
I_{(4,B,+)_n}= \frac{x^{n}y_2^{\frac{n}{2}}(1-x^2 y_2^{-1})}{1-x^4}.
\label{ind-4bn}
\eea 
The index for the representation $(4,B,-)_n$ is
obtained from (\ref{ind-4bn}) by the replacement $y_2\rightarrow y_2^{-1}$.
Only BPS states with $\Delta=0$
contribute to the index\cite{Kinney:2005ej},
and we can also obtain this result by
summing up the contribution from the $\Delta=0$ states in
Table \ref{cps}.

(\ref{ind-4bn}) leads to the index for a vector multiplet on $AdS_4\times \wt S$
\bea
I_{V}= \sum\limits^\infty_{s=0} I_{(4,B,+)_{2s+2}} \chi_s(z'^{2})
=\sum\limits_{P'\in\ZZ} {x^{|P'|+2} \over (1 -x^4)} 
 \frac{(y_2-x^2)y_2^{|P'|}}{1-x^2 y_2} z'^{P'}.
\label{totalindex}
\eea

Let us compare
(\ref{totalindex}) to previous results.

The index computed in \cite{Choi:2008za}
includes only the contribution of the perturbative sector,
which does not include monopole operators.
To compare 
(\ref{totalindex}) with their result,
we need to know which part of the Kaluza-Klein spectrum
corresponds to the perturbative sector.
In other words,
we should know the correspondence between
magnetic charges in the boundary theory and the
charges on the gravity side.
In \cite{Imamura:2009hc} it is shown that
operators without magnetic charges
correspond to Kaluza-Klein modes of the Cartan part
of the vector multiplets with $P=P'=0$.
Therefore, the perturbative part is obtained by picking up the
$P'=0$ term from (\ref{totalindex}) as
\begin{equation}
I_{\rm pert}(x,y_2)=\frac{x^{2}}{1 -x^4}\frac{(y_2-x^2)}{1-x^2 y_2}.
\end{equation}
Because every Cartan $U(1)$ gives the same contributions,
we need to multiply the total rank of the gauge group of the vector multiplets.
In the theory discussed in \cite{Choi:2008za},
$SU(m)$ gauge theory is realized on ${\cal S}$,
and the corresponding index is $(m-1)I_{\rm pert}(x,y_2)$.
Twisted modes on the other locus ${\cal S}'$ belong to $(4,B,-)_n$,
and the corresponding index is $(m-1)I_{\rm pert}(x,y_2^{-1})$.
The total index
$(m-1)I_{\rm pert}(x,y_2)+(m-1)I_{\rm pert}(x,y_2^{-1})$
precisely agrees with (4.15) in \cite{Choi:2008za}.
(We need the replacement $x^2\rightarrow x$, $y_2\rightarrow y$
to match the conventions.)

Twisted sector spectrum including monopole operators is
studied in \cite{Imamura:2009hc}.
The corresponding Kaluza-Klein modes should be obtained from the
spectrum (\ref{dec}) by an appropriate projection.
The analysis in the opposite direction has been already done
in \cite{Imamura:2009hc}.
In the reference, an index for a vector multiplet
in $AdS_4\times \SSS^3$
is conjectured
which reproduces results
on the gauge theory side.
It is (65) in Ref. \cite{Imamura:2009hc}.
In the reference, the chemical potential $y_2$ for the charge $H_2$ is not introduced
and for comparison we should set $y_2$ in (\ref{totalindex}) to $1$.
Then eq.\,(65) in Ref.\,\cite{Imamura:2009hc} perfectly agrees with (\ref{totalindex}).

\section{Summary and discussions}\label{discussions}
In this paper,
we determined the Kaluza-Klein spectrum of
a $d=7$, $\N=2$ vector multiplet in $AdS_4\times \SSS^3$.
After constructing the supersymmetric action of the vector multiplet
in the curved background,
we solved the equation of motion for the scalar fields.
We determined the whole spectrum by using the knowledge of the
representations of the superconformal algebra.
We also performed direct Kaluza-Klein analysis of all the
component fields other than scalar fields in Appendix
\ref{kkanalysis} and obtained the same result.

We computed a character and an index for the Kaluza-Klein spectrum,
and confirmed that
they reproduce the previous results obtained
from
the twisted sectors for $\N=4$ Chern-Simons theories.
The contribution of monopole operators
are correctly reproduced.
This fact
provides an evidence for the correspondence 
between monopole operators in the twisted sector 
and wrapped M2-branes in the dual geometry.

In general,
it is necessary to choose appropriate boundary conditions
for bulk fields in order to establish the field-operator
correspondence.
In the case discussed in this paper,
for a Kaluza-Klein mode with large $s$,
there exists only one boundary condition
which admits a normalizable mode.
However, for small $s$, we have two acceptable
boundary conditions.
One of such bulk fields is
the Kaluza-Klein mode of $A_M$ with spin $1$ and vanishing
internal charges,
which is shown in the third line in Table \ref{spec}.
This mode belongs to a Betti multiplet\cite{D'Auria:1984vy,Fabbri:1999hw}
corresponding to
baryonic $U(1)$ symmetries in the Chern-Simons theory. 
The boundary condition of the gauge field determines 
whether $U(1)$ symmetry is global or local\cite{Witten:2003ya,Marolf:2006nd}.
We chose the boundary condition which makes the $U(1)$ symmetries local.
In this case, monopole operators arise in the
theory as dynamical objects\cite{Imamura:2008ji}.
Taking the opposite boundary condition
corresponds to the $SU(N)$ gauge symmetries rather than $U(N)$s.
It may be interesting to study the relation between the boundary conditions
for the Betti multiplets and corresponding boundary theories
in more detail by using indices.

It is quite important to understand the
operator spectrum of more general Chern-Simons theories
such as $\N=2$ theories
to understand dynamics of Chern-Simons theories and
establish the dual M2-brane description.
Even though it is in general difficult
to compute the spectrum
on the gauge theory side due to large quantum corrections,
it seems possible to extend the analysis of $\N=4$ Chern-Simons theories
to $\N=2$ theories describing M2-branes in orbifold backgrounds.
In such a case, the internal space of the dual geometry in general includes
many two-cycles and has complicated torsion $4$-form cohomology.
The comparison of monopole operator spectrum and Kaluza-Klein
spectrum in such a model would be useful for the identification of
discrete torsion for a given Chern-Simons theory.

\section*{Acknowledgements}
S.Y. would like to thank J.Bhattacharya, S.Kim, S.Minwalla for discussion 
and Tata Institute of Fundamental Research for hospitality. 
Y.I. was supported in part by
Grant-in-Aid for Young Scientists (B) (\#19740122) from the Japan
Ministry of Education, Culture, Sports,
Science and Technology.
S.Y. was supported by
the Global COE Program ``the Physical Sciences Frontier'', MEXT, Japan.

\appendix
\section{Construction of the action}\label{consts}
In this Appendix we construct the ${\cal N}=2$ supersymmetric action
of a vector multiplet in the $7$-dimensional spacetime
$AdS_4 \times \SSS^3$.
We use the Noether procedure with respect to
the background curvature.
We start from the action
\begin{equation}
S_0=\int d^7x\sqrt{-g}\left[
-\frac{1}{4}F_{MN}F^{MN}-\frac{1}{2}\ol\lambda\Gamma^M\nabla_M\lambda-\frac{1}{2}\partial_M\phi_i\partial^M\phi_i\right],
\label{action0}
\end{equation}
and the transformation laws
\begin{equation}
\delta\phi_i=i(\ol\epsilon\sigma_i\lambda),\quad
\delta\lambda=-i\sigma_i\Gamma^M\epsilon\partial_M\phi_i
+\frac{1}{2}F_{MN}\Gamma^{MN}\epsilon,\quad
\delta A_M=-(\ol\epsilon\Gamma_M\lambda),
\label{tr0}
\end{equation}
where $M,N,\ldots$ are seven-dimensional vector indices
and $i,j=1,2,3$ are indices for $SU(2)_R$ triplet.
$\sigma_i$ are Pauli matrices acting on $SU(2)_R$ doublets.
These are obtained by
dimensional reduction from ${\cal N}=1$ supersymmetric Maxwell theory
on the ten-dimensional flat background, and the covariantization
with respect to diffeomorphism.
If the background spacetime is flat and the transformation parameter
$\epsilon$ is a constant,
the action
(\ref{action0}) is invariant under the transformation
(\ref{tr0}).
The first step of the Noether procedure is to compute the supersymmetry variation
without assuming the flatness.
For a general curved background and a coordinate dependent parameter $\epsilon$,
we obtain the variation in the form $J^M\nabla_M\epsilon$.
\begin{equation}
\delta S_0
= \int\left[
i(\ol\lambda\sigma_i\Gamma^M\Gamma^N \nabla_M\epsilon)\partial_N \phi_i
-\frac{1}{2}(\ol\lambda\Gamma^L\Gamma^{MN}\nabla_L\epsilon)F_{MN}
\right].
\label{v1}
\end{equation}
If we were constructing a supergravity action,
this term would be canceled by introducing the Noether coupling
to the gravitino, $J^M\psi_M$.
But now, we want the action invariant under the global supersymmetry,
whose parameter $\epsilon$ satisfies
the Killing spinor equations
\begin{equation}
\nabla_\mu \epsilon = a \Gamma \Gamma_\mu \epsilon, \quad 
\nabla_m \epsilon = b \Gamma \Gamma_m \epsilon, 
\quad
\Gamma=\Gamma_{0123},
\label{nabep}
\end{equation}
where
we use $\mu,\nu,\ldots=0,1,2,3$ for $AdS_4$ and $m,n,\ldots=4,5,6$ for $\SSS^3$.
$a$ and $b$ are parameters with dimension of mass.
These parameters are proportional to the curvature of $AdS_4$ and $\SSS^3$.
By substituting (\ref{nabep}) into
$[\nabla_M, \nabla_N]\epsilon=(1/4)R_{MNPQ} \Gamma^{PQ} \epsilon$
with the curvature
\begin{equation}
R^{\textrm AdS_4}_{\mu\nu\lambda\rho} = - {1\over R_{\textrm AdS_4}^2} (g_{\mu\lambda} g_{\nu\rho} - g_{\nu\lambda} g_{\mu\rho}),\quad
R^{\SSS^3}_{mnpq} = {1\over R_{\SSS^3}^2} (g_{mp} g_{nq} - g_{np} g_{mq})
\label{curvatire}
\end{equation}
we obtain
\begin{equation}
a^2 =\frac{1}{4R_{\textrm AdS_4}^2},\quad
b^2 =\frac{1}{4R_{\SSS^2}^2 }.
\end{equation}
If we require the background spacetime $AdS_4\times \SSS^3$ is the
locus in the M2-brane near horizon geometry $AdS_4\times\SSS^7$,
two radii $R_{AdS}$ and $R_S$ must be related by $R_S=2R_{AdS}$.
This means $a=\pm2b$.
In the following, however, we will not use this relation as an input
because this relation is automatically obtained by requiring
supersymmetry.

Let us first focus on the first term in (\ref{v1}).
By using (\ref{nabep}), we obtain
\begin{equation}
(\mbox{1st term in $\delta S_0$})
=\int\left[
i(2a-3b)(\ol\lambda\sigma_i\Gamma^\nu\Gamma\epsilon) \partial_\nu \phi_i
+i(4a-b) (\ol\lambda\sigma_i\Gamma^n\Gamma\epsilon ) \partial_n \phi_i
\right].
\label{2nd}
\end{equation}
There are two ways to
obtain similar variations to cancel this.
One is to introduce a fermion mass term
\begin{equation}
S'_\lambda= \frac{m_\lambda}{2}(\ol\lambda \Gamma \lambda).
\end{equation}
The second is to deform the fermion transformation law by
\begin{equation}
\delta' \lambda = iq \phi_i\sigma_i \Gamma \epsilon.
\end{equation}
$m_\lambda$ and $q$ are real parameters with mass dimension $1$.
For the variation (\ref{2nd}) to be canceled by
$\delta'S_0$ and $\delta S'_\lambda$,
the parameters $m_\lambda$ and $q$ should be given by
\begin{equation}
m_\lambda=a+b,\quad q=3a-2b.
\end{equation}
In addition to terms canceling (\ref{2nd}),
$\delta' S_0$
and $\delta S'_\lambda$ provide more terms.
One of them is the term in $\delta' S_0$ including $\nabla\epsilon$.
By using the Killing spinor equations in (\ref{nabep})
it becomes
\begin{equation}
i(3a-2b)(4a+3b)(\ol\lambda\sigma_i\epsilon)\phi_i.
\end{equation}
We also obtain a similar term from $\delta' S'_\lambda$:
\begin{equation}
\delta' S'_\lambda
=-i(3a-2b)(a+b)(\ol\lambda\sigma_i\epsilon)\phi_i.
\end{equation}
These two terms are proportional to $\delta\phi_i$,
and can be canceled by introducing the
following scalar mass term:
\begin{equation}
S'_\phi
=-\frac{m_\phi^2}{2}\phi_i\phi_i,\quad
m_\phi^2=-9a^2+4b^2.
\end{equation}

Now all variations independent of the gauge field have been canceled.
Let us turn to the terms including the gauge field.
The second term in (\ref{v1}) is rewritten with
(\ref{nabep}) as
\begin{equation}
(\mbox{$2$nd term in $\delta S_0$})
=\bar\lambda\biggl(
-\frac{3b}{2}F_{\mu\nu} \Gamma^{\mu\nu}
+(2a-b) F_{\nu m}\Gamma^{\nu} \Gamma^m
+\left(2a+\frac{b}{2}\right)F_{mn} \Gamma^{mn}
 \biggr)\Gamma\epsilon.
\label{1st}
\end{equation}
A similar term arises from $\delta S_\lambda'$:
\begin{equation}
\frac{a+b}{2}F_{MN}(\lambda\Gamma\Gamma^{MN}\epsilon).
\label{dslp}
\end{equation}
The first two terms in
(\ref{1st}) must be canceled by the corresponding part of
(\ref{dslp}).
This requires the relation
\begin{equation}
a=2b.
\end{equation}
This is the relation which is expected from the M2-brane near horizon geometry.
The term in (\ref{1st}) including $F_{mn}$ is canceled by introducing
the Chern-Simons term
\begin{equation}
S_{CS}=-\frac{3a}{2}\epsilon^{kmn}A_kF_{mn}.
\end{equation}
Now all variations are canceled, and
we obtain the action \eqref{stot} and the supersymmetry transformation \eqref{susytr} by setting $a=1/(2L)$.

\section{Kaluza-Klein analysis}\label{kkanalysis}
In this Appendix, we carry out Kaluza-Klein analysis
for component fields in a $d=7$, $\N=2$ vector multiplet.
We expand fields in $\SSS^3$ into spherical harmonics,
and determine the conformal dimension for every mode
by using equations of motion derived from the
action we constructed in Appendix \ref{consts}.

We take the Poincare coordinates in $AdS_4$ with the metric
\begin{equation}
ds^2=L^2\frac{-dx_0^2+dx_1^2+dx_2^2+dr^2}{r^2}.
\end{equation}
The conformal dimension is defined as
an eigenvalue of the Lie derivative associated with the
Killing vector
\begin{equation}
D=r\partial_r+x^i\partial_i.
\end{equation}

We denote spin $j$ spherical harmonics in $\SSS^3$
by $Y_{j,(s_1,s_2)}^{m}$.
Let $SU(2)_1\times SU(2)_2$ be the isometry of $\SSS^3$.
The index $j$ is the spin of the field, and quantum numbers $s_1$ and $s_2$
are $SU(2)_1$ and $SU(2)_2$ angular momenta, which takes half integers
satisfying
\begin{equation}
|s_1-s_2|\leq j\leq s_1+s_2,\quad s_1+s_2-j\in\ZZ.
\label{jbound}
\end{equation}
$m$ is the magnetic quantum numbers in the range
\begin{equation}
-j\leq m\leq j.
\end{equation}
$Y_{j,(s_1,s_2)}^{m}$ actually represents a set of $(2s_1+1)(2s_2+1)$ harmonics
forming the $(s_1,s_2)$ representation of $SU(2)_1\times SU(2)_2$.
We suppress indices for them.
Only harmonics with $j=|s_1-s_2|$ are independent.
For example, vector harmonics
$\vec Y_{1,(s,s)}
=(Y_{1,(s,s)}^{+1},Y_{1,(s,s)}^{0},Y_{1,(s,s)}^{-1})$,
which
do not satisfy this condition, are given as the gradient of
the scalar harmonics: $\vec Y_{1,(s,s)}\propto\vec\nabla Y_{0,(s,s)}$.

These harmonics are eigenmodes of the Laplacian:
\begin{equation}
\Delta Y_{j,(s_1,s_2)}^{m}
=-\frac{1}{R^2}[2s_1(s_1+1)+2s_2(s_2+1)-j(j+1)]Y_{j,(s_1,s_2)}^{m},
\label{lap}
\end{equation}
where $R$ is the radius of $\SSS^3$ in which the harmonics are defined.
The following formula is also convenient:
\begin{equation}
\rot Y_{j,(s_1,s_2)}
=\frac{1}{R}[s_2(s_2+1)-s_1(s_1+1)]Y_{j,(s_1,s_2)}.
\label{genrot}
\end{equation}
The differential operator $\rot$ is a generalization of the rotation.
For a spin $j$ field $\phi_j$, 
it is defined by
\begin{equation}
\rot\phi_j=T^{(j)}_m\nabla_m\phi_j,
\label{genrot}
\end{equation}
where $T_m^{(j)}$ are $SO(3)$ generators of spin $j$ representation
normalized by $[T_m^{(j)},T_n^{(j)}]=\epsilon_{mnp}T_p^{(j)}$.
$\rot$ becomes the ordinary rotation
for a vector field, and the Dirac's operator
for a spinor field.
\begin{equation}
\rot\vec\phi_1={\vec\nabla}\times\vec\phi_1,\quad
\rot\phi_{\frac{1}{2}}=-\frac{i}{2}\gamma^m\nabla_m
\phi_{\frac{1}{2}}.
\end{equation}

Kaluza-Klein modes of the scalar fields are studied in the main text.
Let us consider the gauge field.
The linearized equations of motion derived from the action (\ref{stot}) are
\begin{equation}
\nabla_MF^{M\mu}=0,\quad
\nabla_MF^{Mk}-\frac{3}{2L}\epsilon^{kmn}F_{mn}=0.
\end{equation}
The $AdS$ components $A_\mu$ of the gauge field are
scalars in $\SSS^3$ while the $\SSS^3$ components
$A_m$ forms a vector in $\SSS^3$.
They are expanded with scalar and vector spherical harmonics by
\begin{eqnarray}
A_\mu&=&\sum_{s=0}^\infty (a_{s,s})_\mu Y_{0,(s,s)},\\
\vec A&=&
\sum_{s=1}^\infty a_{s-1,s}\vec Y_{1,(s-1,s)}
+\sum_{s=1}^\infty a_{s,s-1}\vec Y_{1,(s,s-1)}
+\sum_{s=1/2}^\infty a_{s,s}\vec\nabla Y_{0,(s,s)}.
\end{eqnarray}
We also expand the gauge transformation parameter $\Lambda$
with the scalar harmonics as
\begin{equation}
\Lambda=\sum_{s=0}^\infty\lambda_{s,s}Y_{0,(s,s)}.
\end{equation}
We can set $a_{s,s}=0$ for $s\geq 1/2$ by using gauge symmetry with parameters $\lambda_{s,s}$
with $s\geq1/2$.
To fix the residual gauge symmetry with parameter $\lambda_{0,0}$
we take the Lorentz gauge in $AdS_4$.
\begin{equation}
\nabla_\mu (a_{0,0})^\mu=0.
\label{gauge2}
\end{equation}
We still have residual gauge symmetry with parameter $\lambda_{0,0}$ satisfying
$\Delta_{AdS_4}\lambda_{0,0}=0$, which will be fixed later.
With the gauge choice we mentioned above,
the equations of motion reduce to
the following set of differential equations.
\begin{equation}
\nabla_\mu(a_{s,s})^\mu=0\quad(s\geq1/2),
\label{eom1}
\end{equation}
\begin{equation}
\left(
\Delta_{AdS_4}(a_{s,s})_\mu
+\frac{3}{L^2}(a_{s,s})_\mu
+(a_{s,s})_\mu\Delta_{\SSS^3}\right)Y_{0,(s,s)}=0,
\label{eom2}
\end{equation}
\begin{equation}
\left(
\Delta_{AdS}a_{s-1,s}
+ a_{s-1,s} \Delta_{\SSS^3}
-\frac{1}{2L^2} a_{s-1,s}
-\frac{3}{L}a_{s-1,s}\vec\nabla\times\right)\vec Y_{1,(s-1,s)}=0,
\label{eom3}
\end{equation}
\begin{equation}
\left(
\Delta_{AdS} a_{s,s-1}
+a_{s,s-1} \Delta_{\SSS^3}
-\frac{1}{2L^2} a_{s,s-1}
-\frac{3}{L}a_{s,s-1}\vec\nabla\times\right)
\vec Y_{1,(s,s-1)}=0.
\label{eom4}
\end{equation}
To determine the conformal dimension
of the corresponding operators, we assume that
$(a_{s,s})_\mu$, $a_{s-1,s}$, and $a_{s,s-1}$ depend only
on the radial coordinate $r$.
By using this assumption and the formuli (\ref{lap}) and (\ref{genrot}),
the equations for
the radial component $(a_{s,s})_r$ become
\begin{equation}
\left(r\frac{d}{dr}-2\right)
(a_{s,s})_r=0,\quad
\left(r^2\frac{d}{dr^2}-2-s(s+1)\right)
(a_{s,s})_r=0.
\end{equation}
For $s\geq1/2$, these two equations do not have non-vanishing solutions.
For $s=0$, there is a solution $(a_{0,0})_r\propto r^2$,
but we can set $(a_{0,0})_r=0$ by
the residual gauge symmetry $\lambda_{0,0}\propto
r^3$, which satisfies $\Delta_{AdS_4}\lambda_{0,0}=0$.
The other equations of motion
(\ref{eom2}), (\ref{eom3}), and (\ref{eom4}) reduce to
\begin{equation}
\left(r^2\frac{d^2}{dr^2}-s(s+1)\right)(a_{s,s})_\mu(r)=0,
\end{equation}
\begin{equation}
\left(r^4\frac{d}{dr}\frac{1}{r^2}\frac{d}{dr}
-s^2-3s\right)a_{s-1,s}(r)=0,
\end{equation}
\begin{equation}
\left(r^4\frac{d}{dr}\frac{1}{r^2}\frac{d}{dr}
-s^2+3s\right)a_{s,s-1}(r)=0.
\end{equation}
Each of these has two independent solutions.
\begin{equation}
(a_{s,s})_\mu\propto r^{s+1},\quad(D=s+2),\quad
r^{-s},\quad(D=-s+1),
\end{equation}
\begin{equation}
a_{s-1,s}\propto r^{s+3},\quad(D=s+3),\quad
r^{-s},\quad(D=-s),
\end{equation}
\begin{equation}
a_{s,s-1}\propto r^s,\quad(D=s),\quad
r^{-s+3},\quad(D=-s+3).
\end{equation}
$D$ given above are the corresponding conformal dimensions.
The former of each equation is the mode given in Table \ref{spec}.

Next, let us consider the fermion field $\lambda$.
The equation of motion is
\begin{equation}
-\Gamma^M \nabla_M\lambda
+\frac{3}{4L}\Gamma\lambda=0.
\label{lambdaeom}
\end{equation}
$\lambda$ is an eight-component spinor off shell and
each mode of this field is expanded by
the direct products of a four-component spinor in
$AdS_4$ and a two-component spinor spherical harmonic in $\SSS^3$.
We take the anzats
\begin{equation}
\lambda=\sum_s\left(\psi_{s-\frac{1}{2},s}(r)\otimes Y_{\frac{1}{2},(s-\frac{1}{2},s)}
+\psi_{s,s-\frac{1}{2}}(r)\otimes Y_{\frac{1}{2},(s,s-\frac{1}{2})}\right).
\label{fermionexpand}
\end{equation}
The coefficient spinors
$\psi_{s-\frac{1}{2},s}$ and
$\psi_{s,s-\frac{1}{2}}$ have an implicit $SU(2)_R$ index
as well as $\lambda$.
Correspondingly
to the factorization of the spinor (\ref{fermionexpand}),
we factorize the Dirac matrices also as
\begin{equation}
\Gamma^m=\gamma^5\otimes\gamma^m,\quad
\Gamma^\mu=\gamma^\mu\otimes{\bf1}_2,\quad
\Gamma=i\gamma^5\otimes{\bf1}_2.
\label{gamma734}
\end{equation}
By substituting (\ref{fermionexpand}) and 
(\ref{gamma734}) into
(\ref{lambdaeom}), we obtain the differential equations
\begin{equation}
\left(r\frac{d}{dr}-\frac{3}{2}\right)\psi_{s-\frac{1}{2},s}
=i\gamma^{\wh r}\gamma^5\left(s+1\right)\psi_{s-\frac{1}{2},s},
\end{equation}
\begin{equation}
\left(r\frac{d}{dr}-\frac{3}{2}\right)\psi_{s,s-\frac{1}{2}}
=i\gamma^{\wh r}\gamma^5\left(-s+\frac{1}{2}\right)\psi_{s,s-\frac{1}{2}},
\end{equation}
where the index $\wh r$ of $\gamma^{\wh r}$ represents
the local Lorentz index along the radial direction.
The solutions to these equations and the corresponding conformal
dimensions are
\begin{equation}
\psi_{s-\frac{1}{2},s}=
\eta_{s-\frac{1}{2},s}^{(+)}
r^{s+\frac{5}{2}}
+
\eta_{s-\frac{1}{2},s}^{(-)}
r^{-s+\frac{1}{2}},\quad
D=s+\frac{5}{2},\quad -s+\frac{1}{2},
\end{equation}
\begin{equation}
\psi_{s,s-\frac{1}{2}}=
\eta_{s,s-\frac{1}{2}}^{(+)}
r^{-s+2}
+
\eta_{s,s-\frac{1}{2}}^{(-)}
r^{s+1},\quad
D=-s+2,\quad s+1,
\end{equation}
where $\eta^{(\pm)}$ are constant spinors
satisfying
$i\gamma^{\wh r}\gamma^5\eta^{(\pm)}=\pm\eta^{(\pm)}$.
$\eta_{s-\frac{1}{2},s}^{(+)}$ and
$\eta_{s,s-\frac{1}{2}}^{(-)}$ correspond to the
modes given in Table \ref{spec}.

\bibliographystyle{utphys}
\bibliography{ref-m2}

\providecommand{\href}[2]{#2}\begingroup\raggedright\begin{thebibliography}{10}

\bibitem{Bagger:2006sk}
J.~Bagger and N.~Lambert, ``{Modeling multiple M2's},'' {\em Phys. Rev.} {\bf
  D75} (2007) 045020,
\href{http://arXiv.org/abs/hep-th/0611108}{{\tt hep-th/0611108}}.

\bibitem{Bagger:2007jr}
J.~Bagger and N.~Lambert, ``{Gauge Symmetry and Supersymmetry of Multiple
  M2-Branes},'' {\em Phys. Rev.} {\bf D77} (2008) 065008,
\href{http://arXiv.org/abs/0711.0955}{{\tt 0711.0955}}.

\bibitem{Bagger:2007vi}
J.~Bagger and N.~Lambert, ``{Comments On Multiple M2-branes},'' {\em JHEP} {\bf
  02} (2008) 105,
\href{http://arXiv.org/abs/0712.3738}{{\tt 0712.3738}}.

\bibitem{Gustavsson:2007vu}
A.~Gustavsson, ``{Algebraic structures on parallel M2-branes},''
\href{http://arXiv.org/abs/0709.1260}{{\tt 0709.1260}}.

\bibitem{Gustavsson:2008dy}
A.~Gustavsson, ``{Selfdual strings and loop space Nahm equations},'' {\em JHEP}
  {\bf 04} (2008) 083,
\href{http://arXiv.org/abs/0802.3456}{{\tt 0802.3456}}.

\bibitem{Aharony:2008ug}
O.~Aharony, O.~Bergman, D.~L. Jafferis, and J.~Maldacena, ``{N=6 superconformal
  Chern-Simons-matter theories, M2-branes and their gravity duals},'' {\em
  JHEP} {\bf 10} (2008) 091,
\href{http://arXiv.org/abs/0806.1218}{{\tt 0806.1218}}.

\bibitem{Bhattacharya:2008bja}
J.~Bhattacharya and S.~Minwalla, ``{Superconformal Indices for ${\cal N}=6$
  Chern Simons Theories},'' {\em JHEP} {\bf 01} (2009) 014,
\href{http://arXiv.org/abs/0806.3251}{{\tt 0806.3251}}.

\bibitem{Kim:2009wb}
S.~Kim, ``{The complete superconformal index for N=6 Chern-Simons theory},''
  {\em Nucl. Phys.} {\bf B821} (2009) 241--284,
\href{http://arXiv.org/abs/0903.4172}{{\tt 0903.4172}}.

\bibitem{Bhattacharya:2008zy}
J.~Bhattacharya, S.~Bhattacharyya, S.~Minwalla, and S.~Raju, ``{Indices for
  Superconformal Field Theories in 3,5 and 6 Dimensions},'' {\em JHEP} {\bf 02}
  (2008) 064,
\href{http://arXiv.org/abs/0801.1435}{{\tt 0801.1435}}.

\bibitem{Hosomichi:2008jd}
K.~Hosomichi, K.-M. Lee, S.~Lee, S.~Lee, and J.~Park, ``{N=4 Superconformal
  Chern-Simons Theories with Hyper and Twisted Hyper Multiplets},'' {\em JHEP}
  {\bf 07} (2008) 091,
\href{http://arXiv.org/abs/0805.3662}{{\tt 0805.3662}}.

\bibitem{Benna:2008zy}
M.~Benna, I.~Klebanov, T.~Klose, and M.~Smedback, ``{Superconformal
  Chern-Simons Theories and $AdS_4/CFT_3$ Correspondence},'' {\em JHEP} {\bf
  09} (2008) 072,
\href{http://arXiv.org/abs/0806.1519}{{\tt 0806.1519}}.

\bibitem{Imamura:2008nn}
Y.~Imamura and K.~Kimura, ``{On the moduli space of elliptic
  Maxwell-Chern-Simons theories},'' {\em Prog. Theor. Phys.} {\bf 120} (2008)
  509--523,
\href{http://arXiv.org/abs/0806.3727}{{\tt 0806.3727}}.

\bibitem{Aharony:1998xz}
O.~Aharony, A.~Fayyazuddin, and J.~M. Maldacena, ``{The large N limit of N =
  2,1 field theories from three- branes in F-theory},'' {\em JHEP} {\bf 07}
  (1998) 013,
\href{http://arXiv.org/abs/hep-th/9806159}{{\tt hep-th/9806159}}.

\bibitem{Gukov:1998kk}
S.~Gukov, ``{Comments on N = 2 AdS orbifolds},'' {\em Phys. Lett.} {\bf B439}
  (1998) 23--28,
\href{http://arXiv.org/abs/hep-th/9806180}{{\tt hep-th/9806180}}.

\bibitem{Choi:2008za}
J.~Choi, S.~Lee, and J.~Song, ``{Superconformal Indices for Orbifold
  Chern-Simons Theories},'' {\em JHEP} {\bf 03} (2009) 099,
\href{http://arXiv.org/abs/0811.2855}{{\tt 0811.2855}}.

\bibitem{Imamura:2009hc}
Y.~Imamura and S.~Yokoyama, ``{A Monopole Index for N=4 Chern-Simons
  Theories},'' {\em Nucl. Phys.} {\bf B827} (2010) 183--216,
\href{http://arXiv.org/abs/0908.0988}{{\tt 0908.0988}}.

\bibitem{Imamura:2008ji}
Y.~Imamura and S.~Yokoyama, ``{N=4 Chern-Simons theories and wrapped M5-branes
  in their gravity duals},''
\href{http://arXiv.org/abs/0812.1331}{{\tt 0812.1331}}.

\bibitem{Imamura:2009ur}
Y.~Imamura, ``{Monopole operators in N=4 Chern-Simons theories and wrapped
  M2-branes},'' {\em Prog. Theor. Phys.} {\bf 121} (2009) 1173--1187,
\href{http://arXiv.org/abs/0902.4173}{{\tt 0902.4173}}.

\bibitem{Witten:2009xu}
E.~Witten, ``{Branes, Instantons, And Taub-NUT Spaces},'' {\em JHEP} {\bf 06}
  (2009) 067,
\href{http://arXiv.org/abs/0902.0948}{{\tt 0902.0948}}.

\bibitem{Klebanov:2010tj}
I.~R. Klebanov, S.~S. Pufu, and T.~Tesileanu, ``{Membranes with Topological
  Charge and AdS4/CFT3 Correspondence},'' {\em Phys. Rev.} {\bf D81} (2010)
  125011,
\href{http://arXiv.org/abs/1004.0413}{{\tt 1004.0413}}.

\bibitem{Benishti:2010jn}
N.~Benishti, D.~Rodriguez-Gomez, and J.~Sparks, ``{Baryonic symmetries and M5
  branes in the $AdS_4/CFT_3$ correspondence},'' {\em JHEP} {\bf 07} (2010)
  024,
\href{http://arXiv.org/abs/1004.2045}{{\tt 1004.2045}}.

\bibitem{Gutierrez:2010bb}
N.~Gutierrez, Y.~Lozano, and D.~Rodriguez-Gomez, ``{Charged particle-like
  branes in ABJM},''
\href{http://arXiv.org/abs/1004.2826}{{\tt 1004.2826}}.

\bibitem{Terashima:2008ba}
S.~Terashima and F.~Yagi, ``{Orbifolding the Membrane Action},'' {\em JHEP}
  {\bf 12} (2008) 041,
\href{http://arXiv.org/abs/0807.0368}{{\tt 0807.0368}}.

\bibitem{Dolan:2008vc}
F.~A. Dolan, ``{On Superconformal Characters and Partition Functions in Three
  Dimensions},''
\href{http://arXiv.org/abs/0811.2740}{{\tt 0811.2740}}.

\bibitem{Kinney:2005ej}
J.~Kinney, J.~M. Maldacena, S.~Minwalla, and S.~Raju, ``{An index for 4
  dimensional super conformal theories},'' {\em Commun. Math. Phys.} {\bf 275}
  (2007) 209--254,
\href{http://arXiv.org/abs/hep-th/0510251}{{\tt hep-th/0510251}}.

\bibitem{D'Auria:1984vy}
R.~D'Auria and P.~Fre, ``{Universal Bose-Fermi mass relations in Kaluza-Klein
  supergravity and harmonic analysis on coset manifolds with Killing
  spinors},'' {\em Ann. Phys.} {\bf 162} (1985)
372.

\bibitem{Fabbri:1999hw}
D.~Fabbri {\em et al.}, ``{3D superconformal theories from Sasakian
  seven-manifolds: New nontrivial evidences for AdS(4)/CFT(3)},'' {\em Nucl.
  Phys.} {\bf B577} (2000) 547--608,
\href{http://arXiv.org/abs/hep-th/9907219}{{\tt hep-th/9907219}}.

\bibitem{Witten:2003ya}
E.~Witten, ``{SL(2,Z) action on three-dimensional conformal field theories with
  Abelian symmetry},''
\href{http://arXiv.org/abs/hep-th/0307041}{{\tt hep-th/0307041}}.

\bibitem{Marolf:2006nd}
D.~Marolf and S.~F. Ross, ``{Boundary conditions and new dualities: Vector
  fields in AdS/CFT},'' {\em JHEP} {\bf 11} (2006) 085,
\href{http://arXiv.org/abs/hep-th/0606113}{{\tt hep-th/0606113}}.

\end{thebibliography}\endgroup

\end{document}